# Speed Radar using HC-SR04 and Arduino: Calibration and Data Analysis

Allan Zapata [1]

[1] EC-106-14927 allan.zapata@up.ac.pa

## 1. Abstract

Ultrasonic sensors are devices that use sound waves to measure distances. They are useful in various applications, such as robotics or factory automation. In this report, the use of these devices for the study of kinematics is presented, specifically to measure the speed of a remote control toy car. The calibration of the sensor achieving a mean absolute error (MAE) of 0,40 cm is presented and then the use of this sensor to measure the speed of the car using an Arduino. The methods of data collection and data analysis are also presented. It is concluded that the ultrasonic sensor is a useful tool for measuring distances and speeds, although its accuracy can be affected by factors such as temperature and humidity.

Los sensores ultrasónicos son dispositivos que utilizan ondas de sonido para medir distancias. Estos son útiles en diversas aplicaciones, como la robótica o automatización en fábricas. En este informe, se presenta el uso de estos dispositivos para el estudio de cinemática, específicamente para medir la rapidez de un auto a control remoto de juguete. Se presenta la calibración del sensor consiguiendo un error absoluto medio (EAM) de 0,40 cm y posteriormente el uso de este para medir la rapidez del auto utilizando un Arduino. Se presenta también los métodos de recolección de datos y análisis de los mismos. Se concluye que el sensor ultrasónico es una herramienta útil para medir distancias y velocidades, aunque su precisión puede verse afectada por factores como la temperatura y la humedad.

**Palabras clave:** HC-SR04; Arduino; Calibración; Datos; Radar de Rapidez; Sensor de Distancia; Física Experimental

## 2. Introducción

El sensor ultrasónico HC-SR04 es un dispositivo ampliamente utilizado en aplicaciones de medición de distancia y detección de objetos [1], [2]. Su funcionamiento depende de dos transductores, un emisor de ondas ultrasónicas —40 kHz— y un receptor del rebote de estas ondas [3]. Midiendo el tiempo $t$ entre la emisión y la recepción de las ondas se puede calcular la distancia $d$ al objeto utilizando la fórmula $d = v_s t$, donde $v_s$ es la velocidad del sonido en el aire. También se puede calibrar el sensor utilizando regresión lineal a partir de mediciones de tiempo de recepción junto con medidas conocidas, lo que permite obtener una relación más precisa entre la distancia medida y la distancia real. En este informe se utiliza este último método.

La decisión de utilizar este método se fundamenta en la cantidad de factores que pueden afectar la precisión de la medición, como la temperatura, densidad y la humedad del aire [3], [4].

## 3. Procedimiento

La realización de este experimento requiere de tres partes: la confección del dispositivo de medición, la calibración del sensor y la medición de la rapidez del auto a control remoto.

### 3.1. Confección del dispositivo de medición

La confección del dispositivo de medición consiste en unir el sensor HC-SR04 a un Arduino MKR WiFi 1010 utilizando un *protoboard* y cables. Este controlador se conectaba a una computadora para servir como fuente de poder y permitir la recolección de datos. Imágenes de los componentes y el dispositivo de medición se pueden ver en la Figura 1.

El sensor ultrasónico HC-SR04 tiene cuatro pines: `Vcc`, `Gnd`, `Trig` y `Echo` *(Véase la Figura 1b)* [3]. El pin `Vcc` se conecta a la fuente de alimentación de 5 V. El pin `Gnd` se conecta a tierra. El pin `Trig`, utilizado para emitir ondas ultrasónicas, se conecta al pin digital 4 del Arduino. El pin `Echo`, utilizado para recibir estas ondas ultrasónicas, se conecta al pin digital 5 del Arduino. El Arduino se alimenta a través de un cable USB conectado a la computadora.

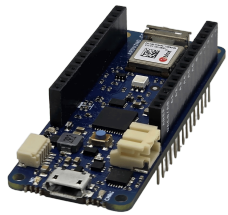 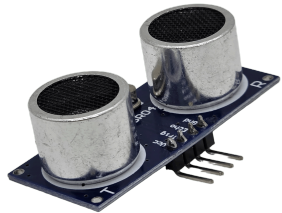 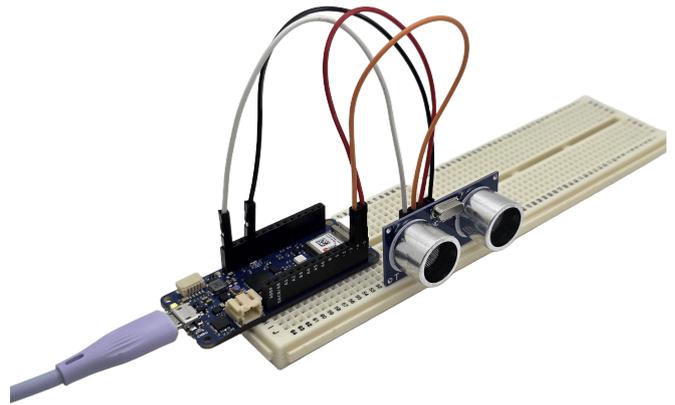

(a) Arduino MKR WiFi 1010    (b) Sensor ultrasónico HC-SR04    (c) Dispositivo de medición

Figura 1: Dispositivo de medición y sus componentes.

Para programar el Arduino se decidió utilizar Platformio[1] ya que este ecosistema permite mayor control que sistemas como Arduino IDE y además permite utilizar sus librerías y arquitecturas. El código en C++ dentro del Arduino se presenta en el Programa 1. Este código incluye la configuración de los pines a utilizar, el uso del transductor emisor y receptor para obtener una lectura y la conversión a distancia utilizando la Ecuación 1 obtenida en la Sección 3.2.

```cpp
#include <Arduino.h>
#include "constants.h"

unsigned long duration;
unsigned long startTime;

void setup() {
  Serial.begin(BAUD_RATE);
  pinMode(TRIG_PIN, OUTPUT);
  pinMode(ECHO_PIN, INPUT);

  startTime = micros();
}

void loop() {
  digitalWrite(TRIG_PIN, LOW);
  delay(2);

  digitalWrite(TRIG_PIN, HIGH);
  delay(10);

  digitalWrite(TRIG_PIN, LOW);

  duration = pulseIn(ECHO_PIN, HIGH);
  Serial.print(duration * 0.0183 - 0.3639);
  Serial.print(" cm - ");
  Serial.print(micros() - startTime);
  Serial.println(" µs");
}
```

Programa 1: Código utilizado en el Arduino para controlar el sensor y registrar sus lecturas.

### 3.2. Calibración del sensor

La calibración del sensor HC-SR04 involucró obtener medidas de los tiempos leídos por este mismo en base a la distancia de un objeto colocado en frente del dispositivo de medición. Se utilizó una regla y una superficie metálica plana para obtener estas medidas *(véase la Figura 2)*.

En la configuración mostrada en la Figura 2, las distancias se empezaban a medir desde la marca de 5 cm en la regla mostrada, esto debido a que es desde donde salen y entran las ondas ultrasónicas del sensor. Algo muy importante a considerar es que no es el punto donde se emiten o se reciben las ondas, puesto que eso ocurre en elementos dentro del sensor que no podemos referenciar.

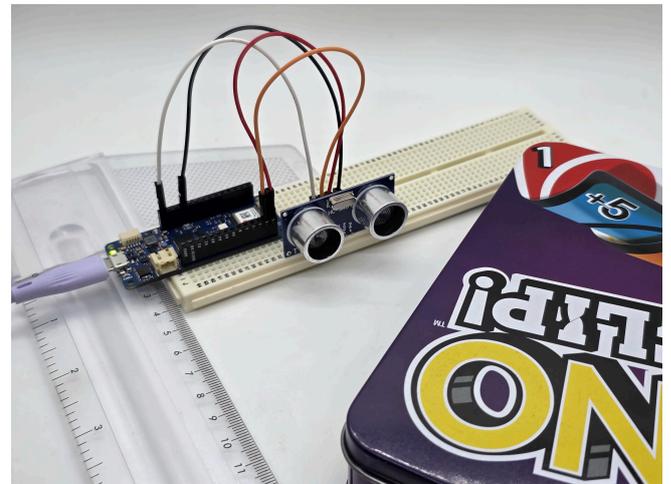

Figura 2: Elementos utilizados en la calibración del sensor

Se obtuvieron 33 medidas al momento de calibrar desde 0,80 cm hasta 29,80 cm. Para cada distancia medida, se colocaba la superficie metálica frente al sensor y se observaba los tiempos resultantes en el monitor serial del Arduino y se registraba el tiempo más frecuente que apareciera en la pantalla. En caso de haber ruido en la medición, lo cual era común, se obtenían 10 medidas en un momento al azar y se calculaba un promedio a partir de esas medidas y ese sería el valor registrado. Esto se repetía 3 veces removiendo y volviendo a colocar la superficie para obtener un valor promedio que sería agregado a la tabla de calibración presentada en la Tabla 1.

---

[1]https://platformio.org/

A partir de esta tabla se realizó una regresión lineal simple para modelar la distancia $L$ en función del tiempo $t$. En este análisis se consideró el tiempo en microsegundos como la variable independiente y la distancia la variable dependiente. Esta regresión y otros cálculos presentados en este informe se encuentran en el *Notebook*[1] adjuntado a este informe.

| $L$ (cm) | $t$ ($\mu$s) | $L$ (cm) | $t$ ($\mu$s) | $L$ (cm) | $t$ ($\mu$s) |
|---|---|---|---|---|---|
| 0.80 | 97.00 | 1.80 | 116.00 | 2.80 | 135.00 |
| 3.80 | 154.00 | 4.30 | 211.00 | 4.80 | 267.00 |
| 5.30 | 305.00 | 5.80 | 320.00 | 6.30 | 360.00 |
| 6.80 | 419.00 | 7.80 | 455.00 | 8.80 | 495.00 |
| 9.80 | 570.00 | 10.80 | 630.00 | 11.80 | 690.00 |
| 12.80 | 753.00 | 13.80 | 826.00 | 14.80 | 846.00 |
| 15.80 | 906.00 | 16.80 | 943.00 | 17.80 | 1018.00 |
| 18.80 | 1055.00 | 19.80 | 1112.00 | 20.80 | 1166.00 |
| 21.80 | 1223.00 | 22.80 | 1260.00 | 23.80 | 1310.00 |
| 24.80 | 1365.00 | 25.80 | 1410.00 | 26.80 | 1460.00 |
| 27.80 | 1500.00 | 28.80 | 1560.00 | 29.80 | 1630.00 |

Tabla 1: Tiempos promedio $t$ en microsegundos obtenidos a partir de las distancias $L$ en el proceso de calibración

La regresión lineal de la distancia sobre el tiempo resultó en la Ecuación 1. Esta regresión tiene un coeficiente de correlación de 0,9985.

$$\lambda = 0.0183t - 0.03639 \quad (1)$$

Con la Ecuación 1 se volvieron a tomar medidas para calcular el error de esta calibración. Se utilizó el mismo procedimiento mencionado anteriormente y se obtuvieron los datos presentados en la Tabla 2

| $L$ (cm) | $\lambda$ (cm) | $L$ (cm) | $\lambda$ (cm) | $L$ (cm) | $\lambda$ (cm) |
|---|---|---|---|---|---|
| 0.50 | 3.15 | 1.00 | 1.58 | 1.50 | 1.99 |
| 2.00 | 1.41 | 2.50 | 1.41 | 3.00 | 3.00 |
| 3.50 | 2.81 | 4.00 | 3.15 | 4.50 | 3.84 |
| 5.00 | 4.18 | 6.00 | 5.45 | 7.00 | 6.53 |
| 8.00 | 7.87 | 9.00 | 9.12 | 10.00 | 9.97 |
| 11.00 | 11.52 | 12.00 | 11.76 | 13.00 | 13.06 |
| 14.00 | 13.92 | 15.00 | 14.78 | 16.00 | 15.63 |
| 17.00 | 17.04 | 18.00 | 17.73 | 19.00 | 18.76 |
| 20.00 | 19.67 | 21.00 | 20.69 | 22.00 | 21.95 |
| 23.00 | 22.75 | 24.00 | 23.73 | 25.00 | 25.12 |
| 26.00 | 26.24 | 27.00 | 27.23 | 28.00 | 27.98 |
| 29.00 | 29.19 | 30.00 | 30.14 | | |

Tabla 2: Distancia real $L$ y distancia leída por el sensor $\lambda$ después de la calibración

---

[1] https://github.com/azapg/velocimetro-hc-sr04/blob/main/velocimetro-hc-sr04.ipynb

Se puede observar en la Tabla 2 que la calibración el sensor registra lecturas muy aproximadas a los valores reales. En el *Notebook* adjunto se hace un pequeño análisis del error de esta calibración y los resultados muestran un EAM de 0,40 cm y un error cuadrático medio (ECM) de 0,37 cm$^2$. Se puede observar en la Figura 3 que hay un mayor error al medir objetos muy próximos al sensor. Se hicieron análisis de error en distintas regiones y se obtuvo que para distancias menores a 10 cm el EAM es de 0,69 cm y para distancias mayores a 20 cm es de 0,18 cm.

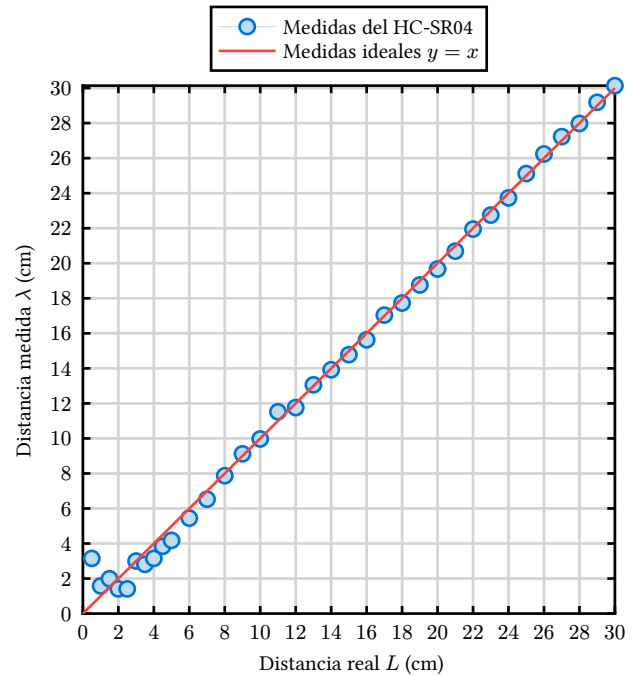

Figura 3: Gráfica mostrando los datos obtenidos en la prueba de la calibración. Se presentan las medidas del sensor y las medidas ideales en base a las medidas reales de las pruebas.

Se intentó hacer una corrección a la Ecuación 1 a partir de la Tabla 2 haciendo una regresión lineal para predecir la distancia real a partir de la medida del sensor, pero se obtuvieron peores resultados. Esta segunda prueba se realizó exclusivamente para distancias mayores a 20 cm y el EAM fue de 0,40 cm, un 118,7 % más que el error sin la corrección. El procedimiento exacto de la corrección se encuentra en el *Notebook* adjunto. Debido a que no se consiguió una mejor calibración que la expresada en la Ecuación 1, fue esta la que se utilizó en las mediciones de rapidez en el Programa 1.

### 3.3. Medición de rapidez

Con el sensor HC-SR04 ya calibrado, se utilizó el código presentado en el Programa 1 para medir la distancia a la que se encontraba el auto de juguete a través del tiempo. En el Programa 1-12 se muestra como se utiliza una medida relativa de tiempo donde se define $t = 0$ en el momento en el que se ejecuta la función `void setup()` en el Arduino.

Utilizando Platformio se guardaban las salidas del monitor serial en un archivo `.log` para el análisis de los datos. Este archivo también agrega una marca de tiempo usando el tiempo real de la computadora.

El auto de juguete se colocaba de frente a 3 cm del sensor. Esto se debe a que el vehículo es más controlable cuando se conduce en reversa. De esta manera es un poco más lento y menos probable a cambiar su dirección inicial.

## 4. Resultados

Después de convertir los archivos `.log` obtenidos del procedimiento presentado en la Sección 3.3 en archivos `.csv` listos para análisis, se obtuvieron los datos de la Figura 4. Se ve claramente como estos datos tienen grandes cantidades de ruido en $1\,\text{s} < t < 2\,\text{s}$. Por otro lado, se puede entender la información general al considerar que el auto a control remoto se encontraba en $x = 3{,}5$ cm en el inicio del experimento y luego acelera hasta que el sensor deja de detectarlo, en $x > 106$ cm.

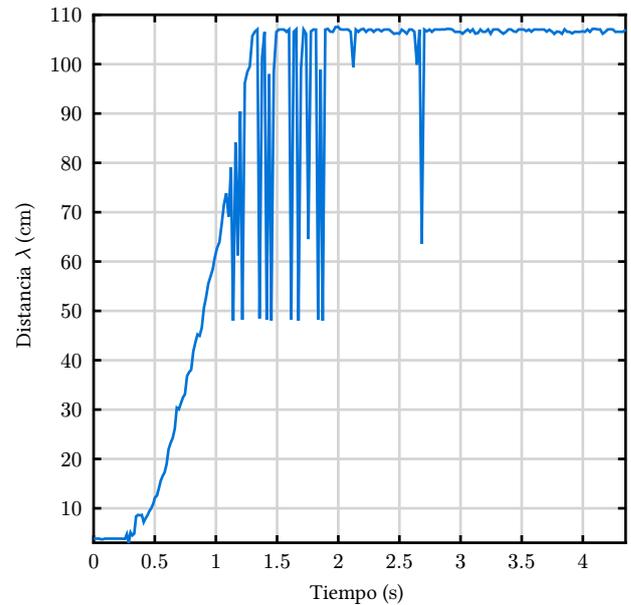

Figura 4: Gráfica que muestra la posición $\lambda$ medida por el sensor en base al tiempo[1]. Solo se muestran los momentos finales del experimento.

Se aplicó un filtro Hampel para eliminar el ruido de medición del conjunto de datos. El filtro utiliza un enfoque de *ventana deslizante* para identificar valores atípicos comparando cada punto con la mediana local en su vecindad. Los parámetros se establecieron con un tamaño de ventana de 13 puntos y un umbral de 0, lo que significa que cualquier punto que se desviara de su mediana local se consideró ruido y se reemplazó con el valor de la mediana. Se eligieron estos parámetros agresivos porque se espera que la señal

---

[1] Se tomó un fragmento de los datos de todos los intentos, realmente, el tiempo empezaría en $t = 610$, pero se transformaron los datos para empezar a contar desde $t = 0$.

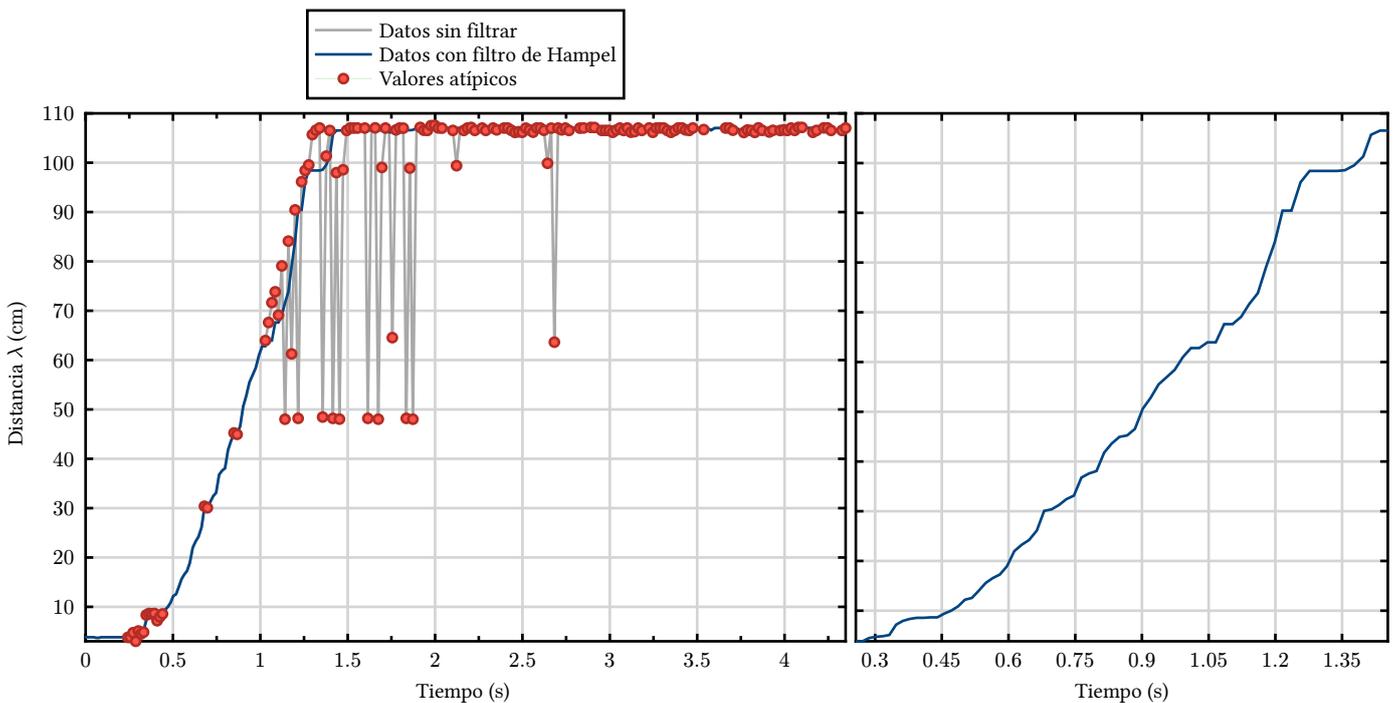

(a) Comparación entre los datos sin filtrar y con filtrado de Hampel.  (b) Movimiento del auto con filtro de Hampel

Figura 5: Gráfica posición $\lambda$ vs tiempo $t$ mostrando la limpieza del ruido utilizando el filtro de Hampel.

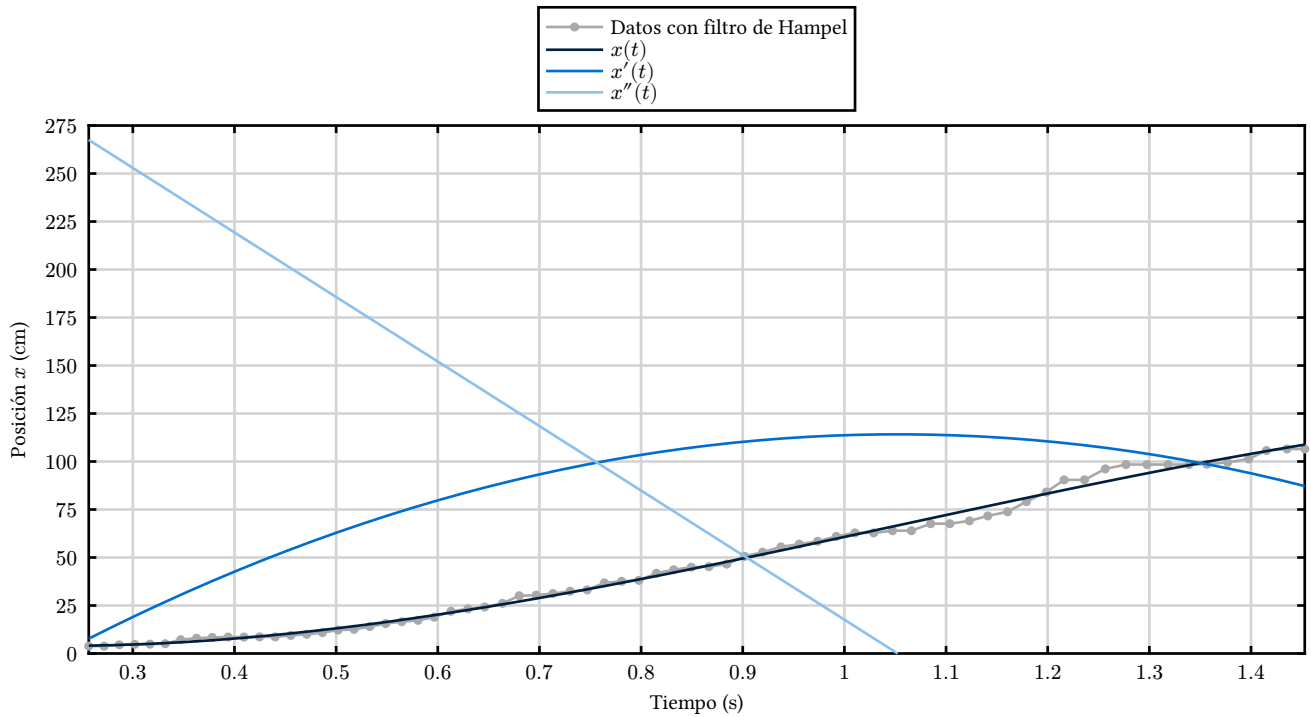

Figura 6: Resultados analíticos del movimiento del auto.

física subyacente (distancia vs. tiempo) sea uniforme y monótona—como una función exponencial—, lo que hace que cualquier desviación pronunciada probablemente se deba a artefactos de medición en lugar de variaciones legítimas de la señal. Los resultados de este filtro se presentan en la Figura 5. El *script* en Python del algoritmo se presenta en el *Notebook* adjunto.

Dadas estas posiciones sin ruido del sensor, se aplicó regresión polinomial de grado tres para obtener una función que describa el movimiento del auto y poder analizar su velocidad y aceleración. El procedimiento para obtener esta regresión se presenta en el *Notebook* adjunto. La regresión dio como resultado el polinomio presentado en la Ecuación 2.[1] Esta regresión se aproxima bastante a los datos, a excepción de las zonas donde se presentaba mayor cantidad de ruido, donde es muy probable que se haya perdido la información original del movimiento del auto. Específicamente, la regresión tiene un EAM de 1,68 cm y un ECM de 5,38 cm$^2$.

$$x = -56.03t^3 + 176.89t^2 - 72.00t + 11.86 \qquad (2)$$

En la Figura 6 se presentan los resultados analíticos obtenidos a partir de la Ecuación 2, incluyendo sus derivadas.

## 5. Discusión

Los resultados muestran ciertos comportamientos contrarios a la intuición, a continuación se discutirán los resultados inesperados sobre ruido y aceleración encontrados en el movimiento.

### 5.1. Resultados analíticos

Inicialmente, se esperaba del análisis que la aceleración del auto fuera positiva en la dirección contraria al sensor, pero la Figura 6 muestra una aceleración negativa, lo cual no es muy intuitivo. Se esperaría que al mover la palanca del control remoto del auto hacia adelante, este empezaría a acelerar positivamente ganando rapidez hasta llegar a un valor constante. Hay varias hipótesis que podrían explicar este resultado, algunas se mencionan a continuación.

**Limitaciones del Motor y Potencia.** El auto a control remoto utilizado es de bajo costo y posee un motor simple. Es probable que este motor genere un impulso inicial considerable (un pico de torque al arrancar), lo que explicaría el aumento inicial de la velocidad. Sin embargo, su potencia sostenida podría ser insuficiente para contrarrestar las fuerzas resistivas que aumentan con la velocidad, como la fricción del aire y la resistencia al rodamiento. Como resultado, el motor podría ser capaz de hacer que el auto alcance rápidamente su velocidad máxima para su potencia limitada bajo esas condiciones, y luego, la aceleración neta se vuelve negativa a medida que las fuerzas de resistencia superan la fuerza de propulsión del motor.

**Respuesta No Lineal del Sistema.** No se puede asumir que el auto, al ser un sistema físico real, responderá de forma perfectamente lineal y continua a la entrada del joystick. Un motor simple, especialmente con una batería económica, puede tener una entrega de potencia que decae o se estabiliza rápidamente, no permitiendo una aceleración positiva sostenida en el tiempo, a pesar de que el usuario mantenga el joystick hacia adelante.

---
[1] Para $\{0{,}26\,\text{s} \leq t \leq 1{,}45\,\text{s}\}$

## 5.2. Ruido

Los datos presentados en la Figura 4 describen la posición medida por el sensor en base al tiempo. La posición inicial del auto era de 3,5 cm, pero el sensor registraba constantemente 3,85 cm. Luego, el auto comenzó a acelerar positivamente hasta salir del área medible del sensor. Este límite se encuentra en $\lambda = 107$ cm. Esto se sabe puesto a que cuando se remueven todos los objetos del sensor, este mide esa cantidad por defecto. Esto se debe a que el experimento fue realizado dentro de la sala de estar del hogar de uno de los participantes, lo que significa que hay muchos objetos (sofás, paredes, sillas) que pueden ser reconocidos por el sensor en vez del auto. En $1\,\text{s} < t < 2\,\text{s}$ de la prueba de velocidad podemos observar grandes cantidades de ruido. Este ruido, generalmente son lecturas incorrectas que regresan a 48,02 cm aunque la media sea mucho mayor que este número. Esto se puede explicar con lo discutido anteriormente: quizás, el sensor reconocía algún objeto cercano cuando sus ondas empezaban a expandirse más dando lecturas falsas.

## 6. Conclusión

El sensor ultrasónico HC-SR04 es una herramienta útil para medir distancias y velocidades, aunque su precisión puede verse afectada por factores como la temperatura y la humedad. En este informe se presentó el uso de este sensor para medir la rapidez de un auto a control remoto de juguete. Se realizó una calibración del sensor obteniendo un error absoluto medio (EAM) de $0{,}40\,\text{cm}$ y posteriormente se utilizó este para medir la rapidez del auto utilizando un Arduino. Se presentaron los métodos de recolección de datos y análisis de los mismos, incluyendo el uso de un filtro Hampel para eliminar el ruido en las mediciones. Se concluyó que el sensor ultrasónico es una herramienta útil para medir distancias y velocidades, aunque su precisión puede verse afectada por factores ambientales, el lugar donde se realiza la medición y las características del objeto a medir. Además, se observó que el auto a control remoto no aceleraba positivamente como se esperaba, lo que podría deberse a limitaciones del motor y la potencia del sistema.